\documentclass[twocolumn,amsmath,amssymb]{revtex4}

\usepackage{graphicx}
\usepackage{color}
\usepackage{mathtools}
\usepackage{multirow}
\begin{document}
\title{Phase Diagram and Snap-Off Transition for Twisted Party Balloons}
\author{Yu-Chuan Cheng$^1$, Ting-Heng Hsieh$^1$, Jih-Chiang Tsai$^2$, and Tzay-Ming Hong$^{1,\ast}$} 
\affiliation{$^1$Department of Physics, National Tsing Hua University, Hsinchu, Taiwan 30013, Republic of China\\
	$^2$Institute of Physics, Academia Sinica, Taipei, Taiwan 11529, Republic of China}
\date{\today}
\begin{abstract}

All children enjoy inflating balloons and twisting them into different shapes and animals. Snapping the balloon into two separate compartments is a necessary step that bears resemblance to the pinch-off phenomenon for water droplet detached from the faucet. In addition to testing whether balloons exhibit the properties of self-similarity and memory effect that are often associated with the latter event, we determine their  phase diagram by experiments. It turns out that a common party balloon does not just snap. They in fact can assume five more shapes, i.e.,  straight, necking, wrinkled, helix, and supercoil, depending on the twist angle and ratio of its length and diameter. Moreover, history also matters due to their prominent hysteresis. One may shift the phase boundary or/and reshuffle the phases by untwisting or lengthening the balloon at different  twist angle and initial length.   Heuristic models are provided to obtain  analytic expressions for the phase boundaries.
\end{abstract}

\maketitle

\begin{figure}
\includegraphics[scale=0.23]{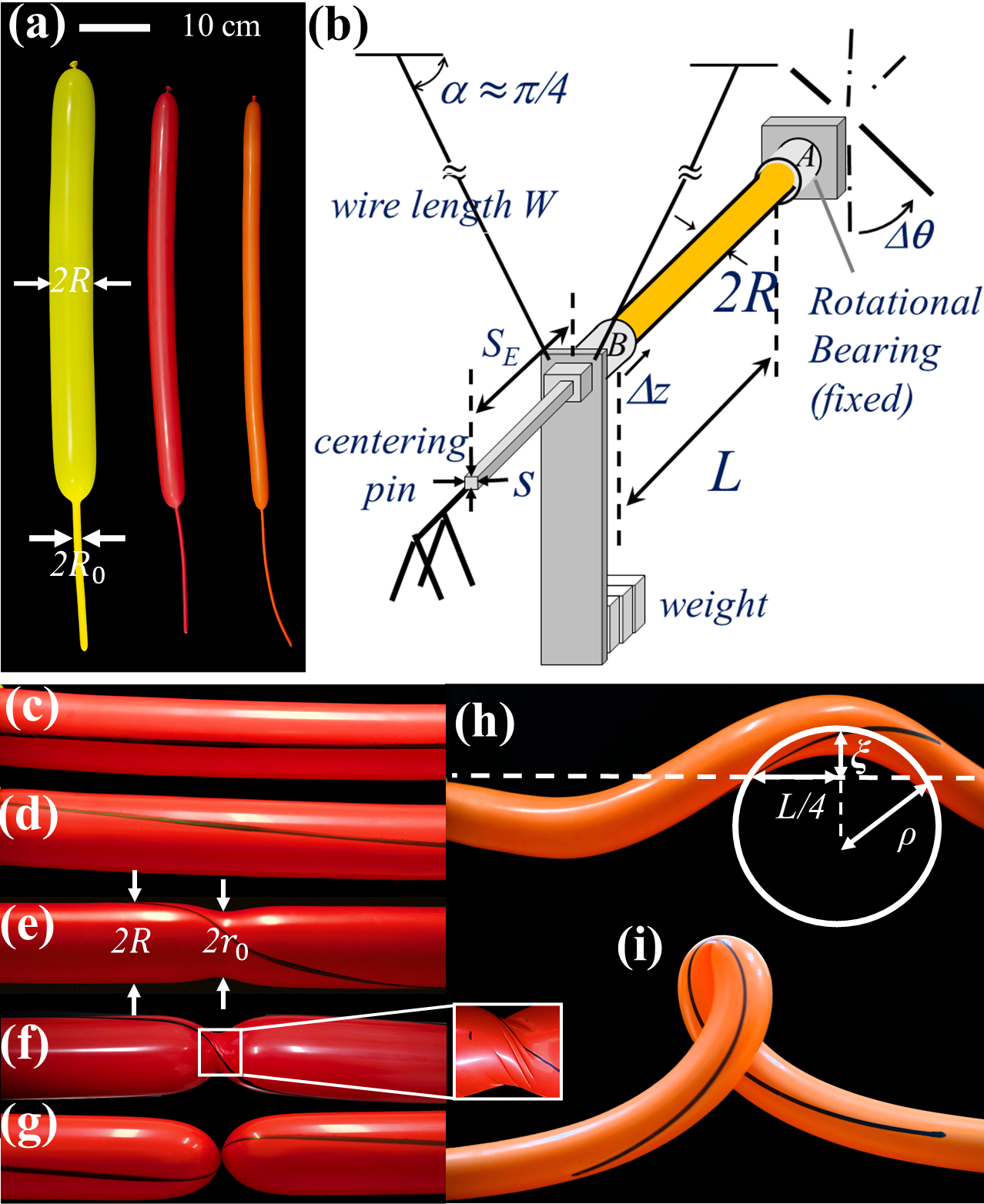} 
\caption{  (a) Sample of inflated balloons with radius $R$ and $R_0$ in the bulged and less-inflated region respectively. (b) Schematic experimental setup. (c) A horizontal mark is drawn before twist. In addition to a straight phase in (d), a medium-sized balloon with $L/2R=160/45$ can develop (e) necking, (f) wrinkles, and (g) snapping.  In contrast, a long balloon with $L/2R=530/45$ shows (h)  helix and (i) supercoil. The white circle in (h) is to facilitate visualization of the radius $\xi$ and $\rho$ of helix and curvature for the helix phase. }
\label{setup}
\end{figure}

Twisting and bending elastic filaments or ribbons \cite{Morigaki:2016:PRL} has fascinated scientists for centuries, such as a coil formed by twisting a rope \cite{Ghatak:2005:PRL} and tendril of the climbing plant \cite{Gerbode:2012:Sci,Goriely:1998:PRL}. In recent years, due to the progress in biology, researches on DNA and protein structures and functions make this topic popular again \cite{Dittmore:2017:PRL,Niew:2019:PRL,Bouchiat:1998:PRL}. For a long rod or wire, a one-dimensional description suffices since the thickness is much smaller than the length. In this Letter, we investigate the twisting of an inflated party balloon, as in Fig. \ref{setup}(a), for which the rubber thickness $t$ is replaced by  radius $R$ that is comparable to length  $L$. As a result, the balloon geometry can be viewed as being intermediate between a 1-D rod \cite{Charles:2019:PRL,Wada:2011:PRE,Ghatak:2007:PRL} and a 2-D ribbon  \cite{Chopin:2013:PRL}.

Balloon is an example of soft material that is easily accessible and fun to play with. When inflated with air or fluid, it exhibits many interesting phenomena, such as a longitudinal phase separation during inflating  \cite{Giudici:2020:PRE}, deformation and bursting when it impacts a rigid wall \cite{Puillet:2020:NP}, fragmentation due to bursting \cite{Moulinet:2015:PRL}, and air transfer between two balloons \cite{Levin:2004:PRE}. Our initial goal  is more mundane, i.e., we want to know  how many configurations a twisted balloon can adopt besides the familiar snapping phase. It turns out that the phase diagram is not a state function of the twist angle $\theta$ and ratio of length and diameter, but also depends on history. For instance, do you crank up $\theta$ or allow it to relax while fixing $L/2R$? Alternatively, one may choose to increase $L/2R$ while fixing $\theta$, in which case the initial value of $L/2R$ and  $\theta$ may also affect the phase boundaries or/and possible configurations.  Although similar hysteretic behavior has been reported for a stretched loop ribbon \cite{Morigaki:2016:PRL}, the rearrangements of soap films in a triangular prism frame \cite{Vandewalle:2011:PRE}, and when the frequency of an applied force to zip or unzip DNA  is varied \cite{Kumar:2013:PRL}, one cannot help but being mesmerized  by clowns who take advantage of this property to transfigure the balloon in a magical way.  Finally, we notice and spend time studying the similarity between the snap-off transition and pinch-off phenomena for water dripping from a faucet \cite{Cohen:1999:PRL, Doshi:2004:PF}, bubble formation \cite{Erlebacher:2011:PRL, Salkin:2016:PRL},  elephant trunks of interstellar gas and dust in the Eagle Nebula \cite{wiki1}, and  the sticky ``capture blob'' used by Bolas spider \cite{wiki2}.

Produced by Sempertex, our balloon samples come in three diameters,  1, 2 and 3 inches. They can be blown up to 60 inches long - about five times their original length. When inflated, their material characteristics are: Young's modulus $Y=0.75$ GPa, thickness $ t=0.07$ mm, and shear modulus $S=0.0002$ GPa.
The experimental setup  in Fig.~\ref{setup}(b) consists mainly of two parts, a frictionless rail and a force sensor. The rail is composed of two hollow coaxial cylinders,  A and B, through which the balloon is inserted.  Cylinder A is the active side via which we twist the balloon around a bearing  fixed on a desk.  As the stepping motor rotates at a steady angular velocity, a tension meter measures the twist force.

 \begin{figure}
\includegraphics[scale=0.4]{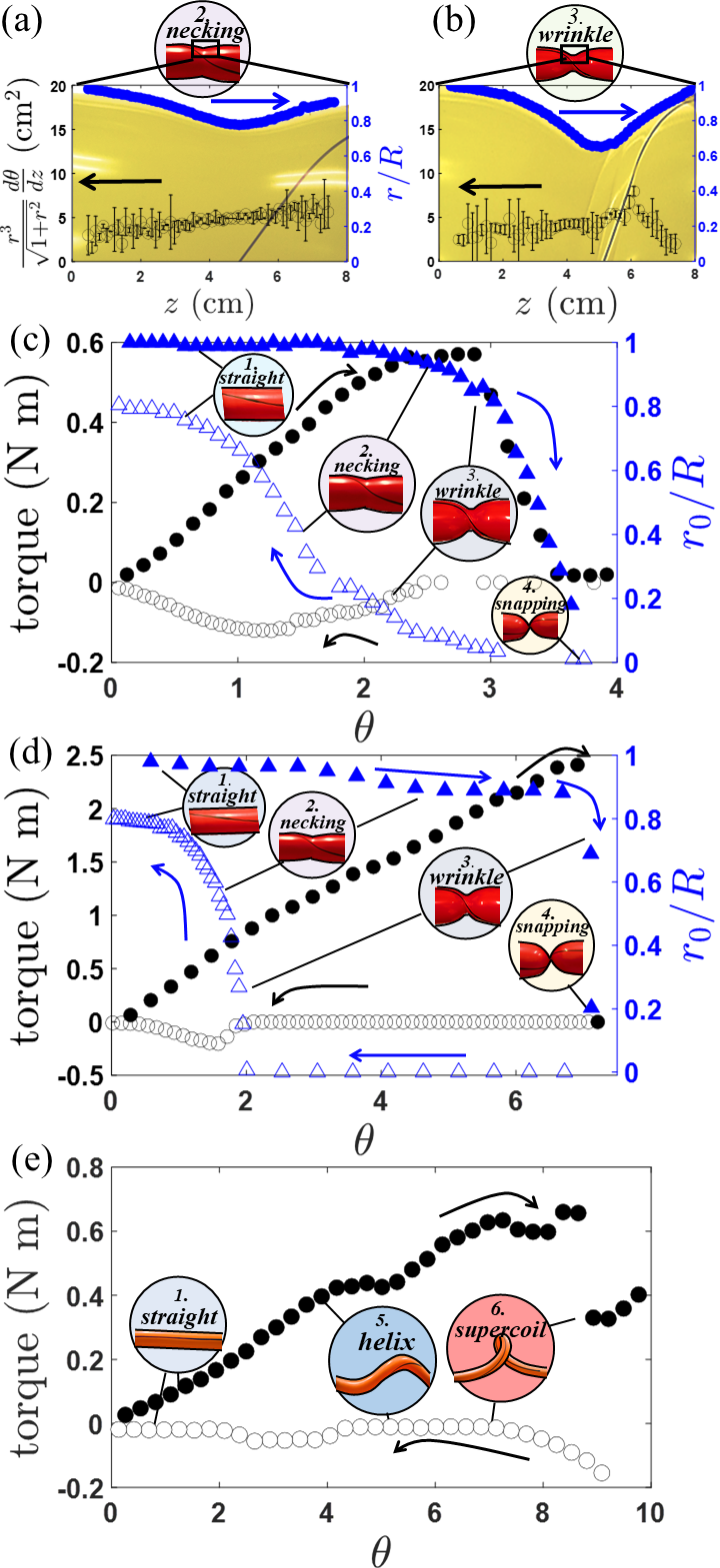} 
\caption
	{ Conserved quantity in Eq. (\ref{torque}) is checked in (a) and (b) for phase-2 and -3 with $L/2R=1.5$. Circle and triangle  represent torque and minimum radius $r_0$, with solid and open  symbols denoting twist and untwist,  in the double-$y$ plot vs $\theta$ for  $L/2R=1.4$ in (c),   2.0 in  (d), and   14 in (e).    }\label{model} 
\end{figure}

Why should the balloon voluntarily concave in phase 2 against the penalty of surface energy? By use of the elasticity relation for shearing, we can rewrite the constant torque in static equilibrium as
\begin{equation}\label{torque}
S\frac{r^3 t }{\sqrt{1+(dr/dz)^2}} \frac{d\theta}{dz}={\rm const.} 
\end{equation}
 This conserved quantity is verified in  Fig. \ref{model} (a,b) for the concave and smooth region. The dip in $4<z<8$ cm in Fig.~\ref{model}(b) is presumably due to the weakening of $S$ by wrinkles. The $r^3$ factor in  Eq. (\ref{torque}) informs us that lowering $r$ can redirect most  shear angle into the concave segment. 

In the following, we refer short  and medium balloons to $L/2R\le 2.5$ and $2.5 < L/2R \le 8$. As shown in Fig. \ref{setup}(d$\sim$f), their configuration evolves continuously from being (1) straight and smooth to (2) concave and smooth and (3) concave and wrinkled. Upon reaching the snapping phase 4 in Fig. \ref{setup}(g),  the balloon suddenly loses its resistance. In contrast to being continuous  in Fig. \ref{model}(c), the  transition in Fig. \ref{model}(d) is abrupt for a medium balloon.
For short balloons,  the torque in Fig. \ref{model}(c) increases linearly with $\theta$ during phases 1 and 2, but bends over in phase 3 before dropping to zero at phase 4. This mechanical response is reminiscent of that for a twisted cylindrical paper or metal shell \cite{limin}, except for the absence of phases 2 and 4 and only a partial region about $2R$ is initially concave.

For a long balloon, the configurations associated with phases (2$\sim$4) are now replaced by (5) helix and (6) supercoil, as shown in Fig. \ref{setup}(h, i). The torque plummets discontinuously with an extent that decreases with each appearance of a new supercoil. A larger slope in its buildup follows this to the next drop. Why? By examining the marker line on the balloon, we know that  shear stress hardly affects the bended region, i.e., the supercoil. Thus, according to Eq. (\ref{e1}) to be shown later, the torque will be inversely proportion to a shortened $L$ that measures only the  straight region. 

Hysteresis is observed when the balloon is allowed to untwist.  The torque in Fig. \ref{model}(c, d) changes continuously from phase 4 to 3 to the best resolution of our force meter, and the neck radius  does not pop up to full $R$ until $\theta$ is much smaller than the critical angle when phase 1 transits to 2 in the forward twist process. This relaxation line is reversible until it meets the forward twisting line that will take over the evolution.  Relaxation for  long balloons  in Fig. \ref{model} (e) is special in that the discontinuous change in torque remains when phase 6 is reverted to 5. Considering that helix and supercoil phases  are shared by twisted filaments, it comes as no surprise that their torque vs. $\theta$ relation is similar \cite{Ghatak:2005:PRL}. Since the marker  line always reverts to horizontal after $\theta$ relaxes to zero, we believe hysteresis is intrinsic and not due to plasticity.

\begin{figure}
\includegraphics[scale=0.17]{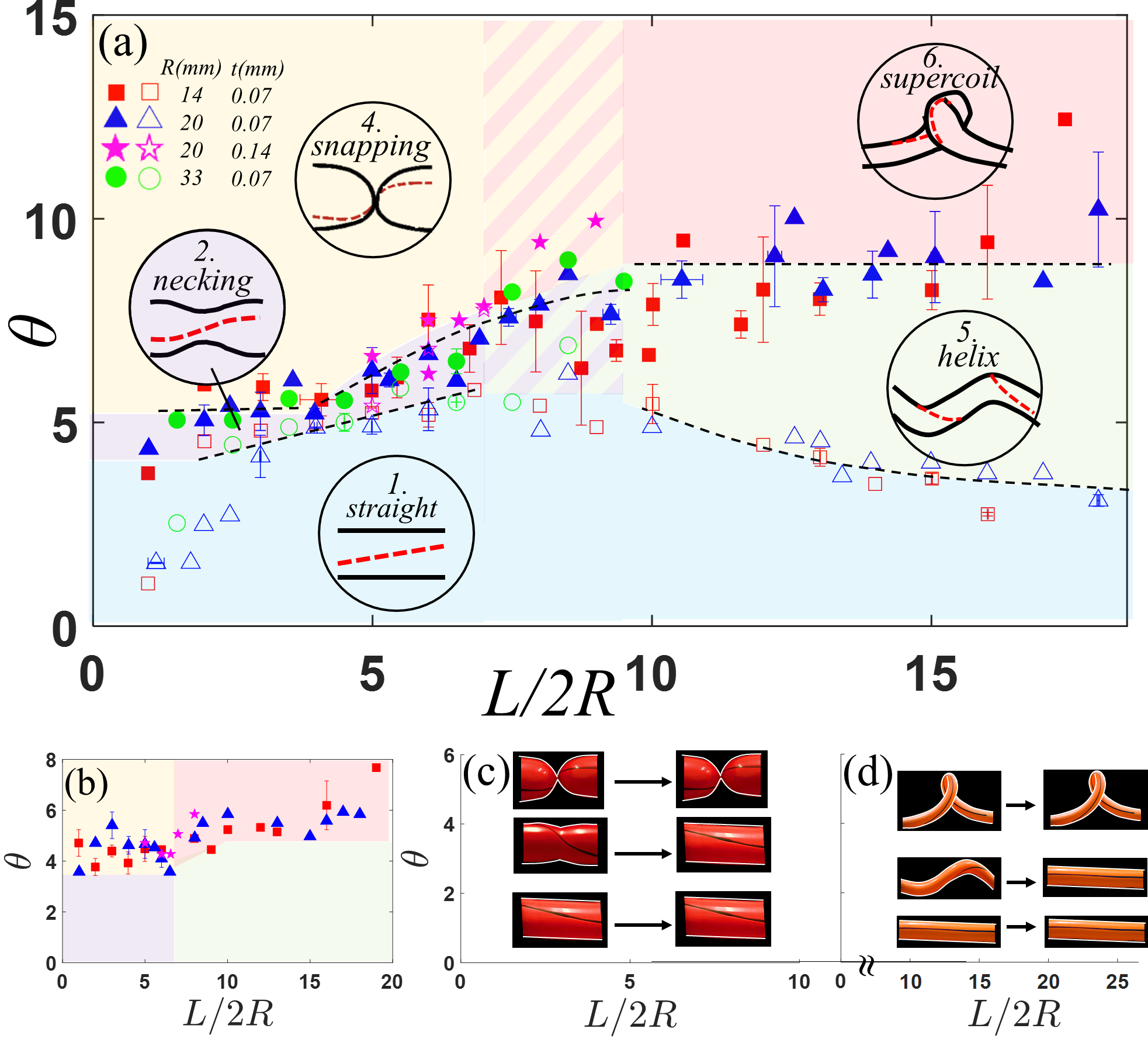}
\caption{Phase diagram  is history-dependent. For instance, (a) is for a twisted balloon with  fixed $L/2R$ where doubling $t$ is achieved by inserting one balloon into another before inflating. Hatch area means there exists a crossover region for the transition. (b) is obtained by untwisting a snap-off or supercoil balloon.  (c) is by lengthening $L$ while fixing $R$ and $\theta$ for a short or medium balloon. Similar to (c), (d) is for a long balloon. Phase 3 is omitted since its range is too small.}
\label{phase_torque}
\end{figure}

In order to determine the phase boundaries in Fig. \ref{phase_torque}(a), we need to write down the potential energy characteristic to all phases. Shear energy for phase 1 is the simplest:
\begin{equation}\label{e1}
E_1 \sim SRtL\Big( \frac{R\theta}{L}\Big)^2=\frac{SR^3t}{L}\theta^2.
\end{equation}
For phase 2, we can approximate the concave segment as being uniform with radius $r\le R$ and obtain
\begin{equation}
E_2\sim \frac{Sr^3t}{\ell}\theta_1^2+\frac{SR^3t}{(L-\ell )}(\theta -\theta_1 )^2
\label{e2}
\end{equation}
Equation (\ref{torque}) requires $r^3\theta_1/\ell =R^3 (\theta-\theta_1 )/(L-\ell )$. The two remaining free parameters can then be determined by optimizing $E_2$ \cite{sm}. Setting $r=R$ and equating Eqs. (\ref{e1}, \ref{e2})  show that the transition from phase 1 to 2 occurs when the shear angle reaches 
\begin{equation}
R\theta_{1,2}/L\sim \sqrt{T/(St)}
\label{theta12}
\end{equation}
 and $\ell\sim R$ where $T$ denotes the surface tension coefficient. Let's skip $E_3$ since its transition from phase 2 is difficult to pin down experimentally. Now it is turn for phase 4 whose main feature  is the creation of two new rubber surfaces about the necking and therefore 
\begin{equation}\label{e4}
E_4 \sim TR^2.
\end{equation}
Physically we expect the snap-off to be triggered by an imbalance between shear and surface energies. In other words, the transition angle $\theta_{3,4}$ can be estimated by equating Eqs. (\ref{e1}) and (\ref{e4}):
\begin{equation}
\theta_{3,4}\sim \sqrt{\frac{T}{St} \frac{L}{R}}.
\label{theta34}
\end{equation}
 
Phase 5 for long balloons is trickier in that  it relieves part of the shear energy by angle $\eta$ by distorting or bending itself into a helix. Furthermore, a second bending term is needed because our experimental setup in Fig.  \ref{setup}(b) requires both ends of the balloon aligned. By twiddling with real balloons, it is easy to realize that this extra energy  (1) is redundant for $\eta =0$ and $2\pi$ when the alignment is automatic, and (2) can be diminished by lengthening the balloon which justifies an additional factor of $R/L$. Overall, we expect
\begin{equation}
E_5 \sim \frac{SR^3t}{L}(\theta-\eta)^2+K_b(\frac{\eta}{L})^2 RL +K_b \frac{\eta (2\pi -\eta )}{L^2} R^2 \\
\label{e5}
\end{equation}
where  $K_b\sim YR^2t$ from elasticity with $\eta/L$ in the second term coming from the curvature $1/\rho$ that can be derived from the Pythagorean theorem in Fig. \ref{setup}(h), $\rho^2=(\rho-\xi)^2+L^2$,   that requires $\rho\sim L^2/\xi$. The fact that $\xi$ is linked to  $\eta$ by  $\xi\sim\eta L$ immediately leads to $\rho\sim L/\eta$. 

Minimizing Eq. (\ref{e5}) with respect to $\eta$ gives
\begin{equation}
\eta \sim \frac{S\theta -Y\pi \frac{R}{L}}{S+Y(1-\frac{R}{L})}. \\
\label{eta}
\end{equation}
In order for phase 1 to enter phase 5 and  $\eta$ to increase from zero, Eq. (\ref{eta}) sets a lower bound on $\theta$ that can be duly identified as the  boundary
\begin{equation}
\theta_{1,5}\sim \frac{Y}{S}\frac{R}{L}. \\
\label{theta15}
\end{equation}
As  the twisting continues, we come to the next boundary  by setting $\eta=2\pi$ in Eq. (\ref{e5}). This renders
\begin{equation}
\theta_{5,6}\sim 2 \pi+ \frac{Y}{S},
\label{theta56}
\end{equation}
independent of $L,R$. The same conclusion has been derived \cite{Goriely} for a twisted filament.

Having established in Fig. \ref{model}(c-e) that the value of torque and $r_0$ can be different at the same $\theta$, depending on the sign of $d\theta/dt$, we thus expect a different phase diagram for relaxation. As plotted in  Fig. \ref{phase_torque}(b), it turns out that the phase boundary of $\theta_{3,4}$ indeed changes. Instead of following  Eq. (\ref{theta34}), its value decreases and is roughly independent of $L/2R$ for short and medium balloons. For long balloons, $\theta_{5,6}$ now increases rather than decreases with $L/2R$ as Eq. (\ref{theta56}). In the mean time, $\theta_{1,2}$ and $\theta_{1,5}$ becomes negligibly small for all lengths, in contrast to Eqs. (\ref{theta12}) and (\ref{theta15}). It is surprising that a  system as simple as the party balloon should exhibit a third or even fourth phase diagram when we lengthen $L$. Experimentally this is achieved by shifting  the moving end B in Fig. \ref{setup}(a)  while holding the balloon to keep the twist angle fixed. If we start from phase 4, it remains so without entering phase 6, as shown in Fig. \ref{phase_torque}(c). But if we start from phase 5, it will transit to phase 1 in Fig. \ref{phase_torque}(d).

\begin{figure}
\includegraphics[scale=0.26]{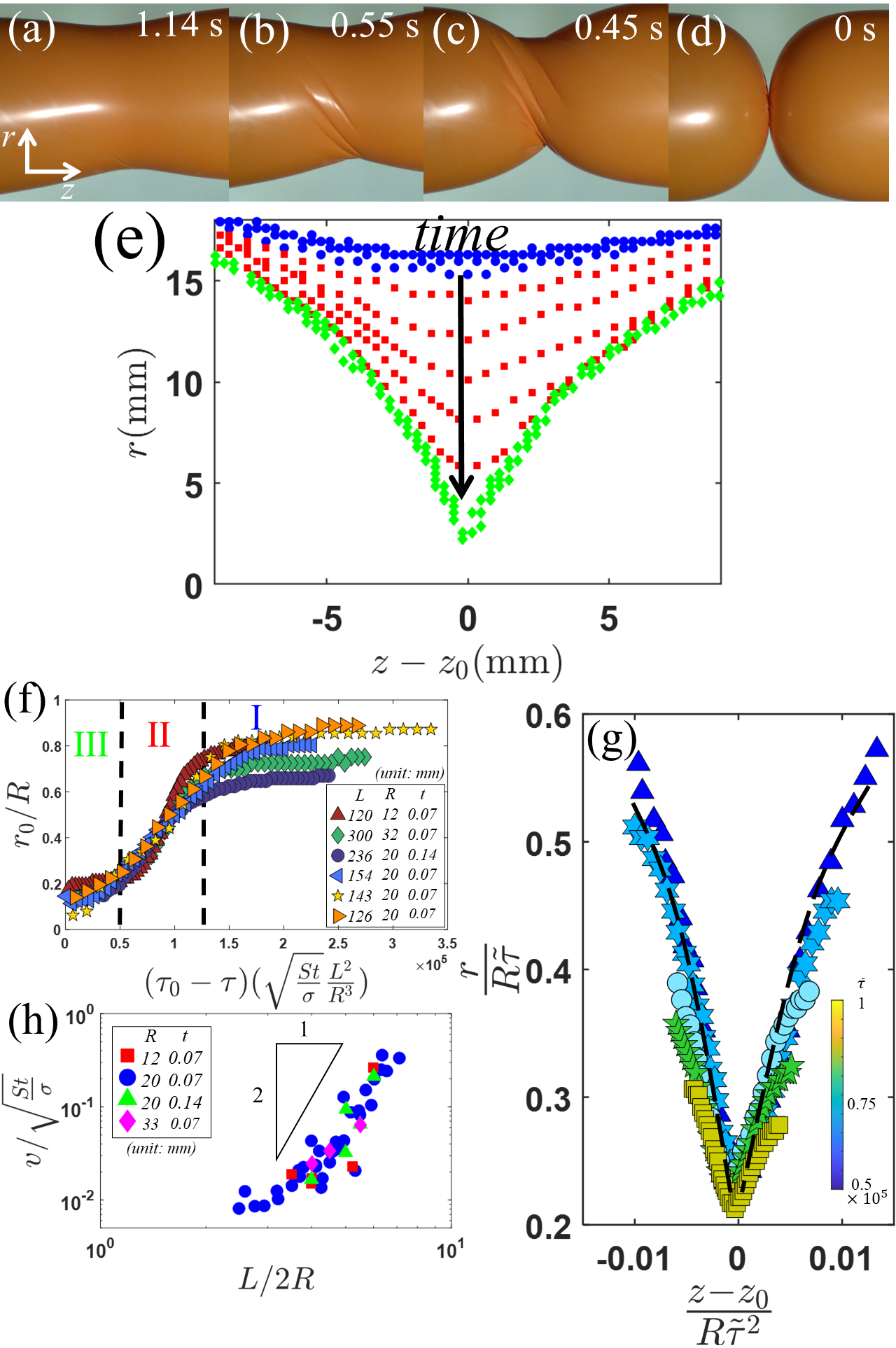} 
\caption{Photos in (a, b, c) are  for different $\theta$ with  $L/2R=3.6$.  The snap-off phase in (d) shares the same angle as (c), but was taken later.  (e) Time evolution of balloon  profile where $z_0$ is the position at the bottle neck. (f) Renormalized $r_0/R$ vs. $\tilde{\tau}$ for balloons of different $L, R,$ and $t$. Dash lines separate three different regimes.  (g) Data  in regime II can be made to collapse to a master curve by rescaling with $r/R\tilde{\tau}$ and $(z-z_0)/R\tilde{\tau}^2$.  (h) Shrinking speed for regime II increases quadratically in response to $L/2R$.  }
\label{fig:evolution}
\end{figure}

Like the collapse of a over-stretched bubble film\cite{weizhi}, we expect the configuration of a medium balloon to transit from  reversible to  irreversible beyond $\theta_{3,4}$. A high-speed camera with 8000 fps captures the dynamical process that leads up to the singular neck. Photos in Fig. \ref{fig:evolution}(a) and (b) are reversible and correspond to phase 2 and 3. Being taken at successive time at the same $\theta$, Fig. \ref{fig:evolution}(c) and (d), categorized as phase 3 and 4,  thus reveals their irreversibility.  In the following, we define $\tilde{\tau}=(\tau_{0}-\tau)(\sqrt{\frac{St}{\sigma}}L^2)/R^3$ where $\sigma$ is the surface density, $\tau_0$ is the time of pinch-off and $\tau$ is real time. Their profile mapped out by photo detection in Fig. \ref{fig:evolution}(e) can be divided into three regimes. Regime I is reversible in which the shrink speed $v$ is determined by  the twist velocity.  After a brief acceleration, $v$ is roughly a constant in most of regime II, as demonstrated in Fig. \ref{fig:evolution}(f) that shows $r_0/R$ vs $\tilde{\tau}$.  This can be understood by comparing the energy scales $\sigma v^2\sim St$ from inertia and shear stress. Deceleration finally halts  the shrinkage in regime III. 

To test whether there exists the property of self-similarity that is characteristic of the pinch-off phenomenon \cite{Cohen:1999:PRL,Aagesen,Lo:2019:PRL,Lister,Eggers,Zhang1999,Lister1998,Rubio2019,Ruth:2019:PNAS,Burton:2004:PRL,Burton:2007:PRL}, we rescale the profiles in regime II by  $r/(R\tilde{\tau}^\alpha)$ and  $(z-z_0)/(R\tilde{\tau}^\beta)$ where $\alpha=1$ and $\beta=2$. As shown in Fig. {\ref{fig:evolution}(g),  they can indeed be made to collapse to a master curve. The scaling exponent $\alpha =1$ is expected from the constant $v$. Why is our observation $\alpha <\beta$ different from  $0<\beta\le \alpha\le 1$ for most pinch-off systems \cite{Cohen:1999:PRL}? For starters, rubber does not obey the Rayleigh-Plateau  instability that predicts the breakage of water column occurs when $r\le z-z_0$. Therefore, $\beta-\alpha<0$ is required to revert the relative size in
\begin{equation}
\frac{r}{\tilde{\tau}^\alpha}>\frac{z-z_0}{\tilde{\tau}^\beta}\tilde{\tau}^{\beta-\alpha}
\end{equation}
at small $\tilde{\tau}$. Figure \ref{fig:evolution}(h) shows that $v\propto(L/2R)^2$, as expected  from Eq. (\ref{theta34}) since longer balloons pump more twist angle and shear energy into the concave segment at the snapping transition \cite{sm}. It is worth noting that this dependence on the initial length violates the lack of memory effect which property is often flaunted for pinch-off systems where  $v$ only depends on the flow of inner liquid \cite{Shi:1994:Sci,Doshi:2003:Sci}. There has been a debate whether this oblivion of initial condition is a natural consequence of the scaling property \cite{Doshi:2003:Sci,Keim:2019:RMP,Pahlavan:2019:PNAS}.  Had our snap-off transition  shared the same physics as pinch-off phenomenon, the fact that Fig. \ref{fig:evolution}(e-h) bleaches this correlation would have supported the negative view held by Ref. \cite{Doshi:2003:Sci}.

In summary, we show that party balloons do not always snap when twisted. It in fact exhibits five other configurations, depending on the twist angle and aspect ratio between length and diameter. Physical arguments, including the finding of a conserved quantity that links radius and the gradient of strain, are provided to derive and verify boundaries  in the phase diagram obtained empirically.  Hysteresis is surprisingly pronounced in this simple system. The allowed phases and location of phase boundaries are subject to change not only when the balloon is untwisted, but when lengthened at different twist angles.
Special attention is paid to clarify the analogy between snapping  and the well-studied pinch-off phenomenon in fluid dynamics. Unlike  the latter, the shrinking speed of  balloon is sensitive to the initial length and radius.

We gratefully acknowledge  financial support from MoST in Taiwan under Grants No. 105-2112-M007-008-MY3 and No. 108-2112-M007-011-MY3.


\begin{thebibliography}{}
	\item[$\ast$] ming@phys.nthu.edu.tw
	\bibitem{Morigaki:2016:PRL} Y. Morigaki, H. Wada, and Y. Tanaka, Phys. Rev. Lett. {\bf 117}, 198003 (2016).
	\bibitem{Ghatak:2005:PRL} A. Ghatak and L. Mahadevan, Phys. Rev. Lett. {\bf 95}, 057801 (2005).
	\bibitem{Gerbode:2012:Sci} S. J. Gerbode, J. R. Puzey, A. G. McCormick, and L. Mahadevan, Science {\bf 337}, 1087 (2012).
	\bibitem{Goriely:1998:PRL} Alain Goriely and Michael Tabor,  Phys. Rev. Lett. {\bf 80}, 1564 (1998).
	\bibitem{Dittmore:2017:PRL} A. Dittmore, S. Brahmachari, Y. Takagi, J. F. Marko, and K. C. Neuman, Phys. Rev. Lett. {\bf 119}, 147801 (2017).
	\bibitem{Niew:2019:PRL} Szymon Niewieczerzal and J. I. Sulkowska,  Phys. Rev. Lett. {\bf 123}, 138102 (2019).
	\bibitem{Bouchiat:1998:PRL} C. Bouchiat and M. Mezard, Phys. Rev. Lett. {\bf 80}, 1556 (1998).
	\bibitem{Charles:2019:PRL} N. Charles, M. Gazzola, and L. Mahadevan, Phys. Rev. Lett. {\bf 123}, 208003 (2019).
	\bibitem{Wada:2011:PRE} Hirofumi Wada, Phys. Rev. E {\bf 84}, 042901 (2011).
	\bibitem{Ghatak:2007:PRL} Animangsu Ghatak and Apurba Lal Das, Phys. Rev. Lett. {\bf 99}, 076101 (2007).
	\bibitem{Chopin:2013:PRL} J. Chopin and A. Kudrolli, Phys. Rev. Lett. {\bf 111}, 174302 (2013).
	\bibitem{Giudici:2020:PRE} A. Giudici and J. S. Biggins, Phys. Rev. E {\bf 102}, 033007 (2020).
	\bibitem{Puillet:2020:NP} E. Jambon-Puillet, T. J. Jones and P.-T. Brun, Nat. Phys. {\bf 16}, 585 (2020).
	\bibitem{Moulinet:2015:PRL} S. Moulinet and M. Adda-Bedia, Phys. Rev. Lett. {\bf 115}, 184301 (2015).
	\bibitem{Levin:2004:PRE} Y. Levin and F. L. da Silveira,  Phys. Rev. E {\bf 69}, 051108 (2004).
	\bibitem{Vandewalle:2011:PRE} N. Vandewalle, M. Noirhomme, J. Schockmel, E. Mersch, G. Lumay, D. Terwagne, and S. Dorbolo, Phys. Rev. E {\bf 83}, 021403 (2011).
	\bibitem{Kumar:2013:PRL} S. Kumar and G. Mishra, Phys. Rev. Lett. {\bf 110}, 258102 (2013).
	\bibitem{Doshi:2004:PF} P. Doshi and O. A. Basaran,  Phys. Fluids {\bf 16}, 585 (2004).
	\bibitem{Cohen:1999:PRL} I. Cohen, M. P. Brenner, J. Eggers, and S. R. Nagel, Phys. Rev. Lett. {\bf 83}, 1147 (1999).
	\bibitem{Erlebacher:2011:PRL} J. Erlebacher,  Phys. Rev. Lett. {\bf 106}, 225504 (2011).
	\bibitem{Salkin:2016:PRL} L. Salkin, A. Schmit, P. Panizza, and L. Courbin,  Phys. Rev. Lett. {\bf 116}, 077801 (2016).
	\bibitem{wiki1} \url{https://en.wikipedia.org/wiki/Pillars_of_Creation.}
	\bibitem{wiki2} \url{https://en.wikipedia.org/wiki/Bolas_spider.}
	\bibitem{limin} L. M. Wang, S. T. Tsai, C. Y. Lee, P. Y. Hsiao, J. W. Deng, H. C. Fan Chiang, Y. Fei, and T. M. Hong, Phys. Rev. E {\bf 101}, 053001 (2020).
	\bibitem{sm} See Supplemental Material at *** for detailed calculations.
	\bibitem{Goriely} A. Goriely and M. Tabor, Nonlinear Dynamics  {\bf 21}, 101 (2000). 
	\bibitem{weizhi} W. C. Li, C. Y. Shih, T. L. Chang, and T. M. Hong (unpublished).
	\bibitem{Lo:2019:PRL} H. Y. Lo, Y. Liu, S. Y. Mak, Z. Xu, Y. Chao, K. J. Li, H. C. Shum, and L. Xu,  Phys. Rev. Lett. {\bf 123}, 134501 (2019).
	\bibitem{Aagesen} L. K. Aagesen, A. E. Johnson, J. L. Fife, P. W. Voorhees, M. J. Miksis, S. O. Poulsen, E. M. Lauridsen, F. Marone, and M. Stampanoni, Nat. Phys. {\bf 6}, 796 (2010).
	\bibitem{Lister} J. R. Lister and H. A. Stone, Phys. Fluids {\bf 10}, 2758 (1998).

	\bibitem{Eggers} J. Eggers,  Phys. Rev. Lett. {\bf 71}, 3458 (1993).
	\bibitem{Zhang1999} W. W. Zhang and J. R. Lister, Phys. Rev. Lett. {\bf 83}, 1151 (1999).
	\bibitem{Lister1998} J. R. Lister and H. A. Stone, Phys. Fluids {\bf 10}, 2758 (1998).
	\bibitem{Rubio2019} M. Rubio, A. Ponce-Torres, E. J. Vega, M. A. Herrada, and J. M. Montanero, Phys. Rev. Fluids {\bf 4}, 021602 (2019).
	\bibitem{Ruth:2019:PNAS} D. J. Ruth, W. Mostert, S. Perrard, and L. Deike, Proc. Natl. Acad. Sci. U.S.A. {\bf 116}, 25412 (2019).
	\bibitem{Burton:2004:PRL} J. C. Burton, J. E. Rutledge, and P. Taborek,  Phys. Rev. Lett. {\bf 92}, 244505 (2004).
	\bibitem{Burton:2007:PRL} J. C. Burton and P. Taborek, Phys. Rev. Lett. {\bf 98}, 224502 (2007).
	\bibitem{Shi:1994:Sci} X. D. Shi, M. P. Brenner, and S. R. Nagel,  Science {\bf 265}, 219 (1994).
	\bibitem{Doshi:2003:Sci} P. Doshi, I. Cohen, W. W. Zhang, M. Siegel, P. Howell, O. A. Basaran, and S. R. Nagel, Science {\bf 302}, 1185 (2003).	
	\bibitem{Keim:2019:RMP} N. C. Keim, J. D. Paulsen, Z. Zeravcic, S. Sastry, and S. R. Nagel, Rev. Mod. Phys. {\bf 91}, 035002 (2019).	
	\bibitem{Pahlavan:2019:PNAS} A. A. Pahlavan, H. A. Stone, G. H. McKinley, and R. Juanes, Proc. Natl. Acad. Sci. U.S.A. {\bf 116}, 13780 (2019).
	




	
	



\end{thebibliography}
\end{document}